\documentclass[aps,prl,twocolumn,superscriptaddress,graphics,showpacs]{revtex4}
\usepackage{graphicx}

\begin{document}

%\preprint{}

\title{Velocity Statistics Distinguish Quantum Turbulence from Classical Turbulence}


\author{M. S. Paoletti}
\affiliation{Departments of Physics, Geology, Institute for Research in Electronics
and Applied Physics, and}
\author{Michael E. Fisher}
\affiliation{Institute for Physical Sciences and Technology, University of Maryland, College Park, MD 20742}
%\thanks{}
\author{K. R. Sreenivasan}
\affiliation{Institute for Physical Sciences and Technology, University of Maryland, College Park, MD 20742}
\affiliation{International Centre for Theoretical Physics, Trieste, Italy 34014}
\author{D. P. Lathrop}
\email[email address: ]{lathrop@umd.edu}
\affiliation{Departments of Physics, Geology, Institute for Research in Electronics
and Applied Physics, and}
\affiliation{Institute for Physical Sciences and Technology, University of Maryland, College Park, MD 20742}

\begin{abstract}
By analyzing trajectories of solid hydrogen tracers, we find that the distributions of velocity in decaying quantum turbulence in superfluid $^4$He are strongly non-Gaussian with $1/v^3$ power-law tails.  These features differ from the near-Gaussian statistics of homogenous and isotropic turbulence of classical fluids.  We examine the dynamics of many events of reconnection between quantized vortices and show by simple scaling arguments that they produce the observed power-law tails.
\end{abstract}

%\keywords{}

\maketitle

The pioneering work of Kolmogorov \cite{kolmogorov41a, kolmogorov41b} remains the cornerstone of the statistical theory of classical incompressible turbulence.  Kolmogorov made two key assumptions: (1) local isotropy and homogeneity prevail, and (2) there exists an inertial range in which turbulent energy is transferred from large to small scales independent of viscosity and generation mechanisms.  Dimensional arguments then yield the spectral density $E(k)=c_K\bar{\epsilon}^{2/3}k^{-5/3}$, where $\bar{\epsilon}$ is the average energy dissipation rate per unit mass and $c_K$ is the universal Kolmogorov constant.  While experiments have found the effects of intermittency to be important for high-order moments, the correction to the spectral form is quite small.  Furthermore, in both experiment \cite{noullez97} and direct numerical simulations \cite{vincent91, gotoh02} the velocity in homogenous and isotropic turbulence is found to exhibit near-Gaussian statistics.

Quantum fluids, however, are typically described as a mixture of two interpenetrating fluids \cite{donnelly91}, a viscous normal fluid and an inviscid superfluid exhibiting long-range quantum order.  There is no conventional viscous dissipation in the superfluid component and vorticity is confined to one-dimensional quantized vortices which possess circulation values that are integer multiples of $\kappa=h/m=9.97\times 10^{-4}$ cm$^2$/s \cite{donnelly91}.  Thus, quantum turbulence takes the form of a complex tangle of atomically-thin vortex filaments of quantized strength \cite{feynman55}.  Dissipation in the superfluid component for 1.70 K $<T<$ 2.05 K is mainly produced by mutual friction \cite{vinen57} between the quantized vortices and normal fluid.

Despite these fundamental differences, there have been notable studies demonstrating similarities between quantum and classical turbulence \cite{samuels92, smith93, barenghi97, nore97, vinen00, maurer98, kobayashi05, stalp99, skrbek03, walmsley07}.  Even though the quality of the supporting evidence has been questioned \cite{procaccia08}, it will be summarized here.  Experiments on turbulence generated in $^4$He by two counter-rotating disks observed Kolmogorov energy spectra that were indistinguishable above and below the superfluid transition \cite{maurer98}.  The Kolmogorov energy spectrum was seen in numerical simulations of the Gross-Pitaevskii equation with small-scale dissipation added to the otherwise energy-conserving dynamics \cite{kobayashi05}.  The classical decay of vorticity \cite{smith93} has been observed in towed grid \cite{smith93, stalp99}, thermal counterflow \cite{skrbek03}, and impulsive spin down \cite{walmsley07} experiments.  In all these studies the flow scales observed were considerably larger than typical intervortex spacings.  These results may be attributed to the fact that on such scales the pairwise interactions of quantized vortices are insignificant while the normal and superfluid components become \lq\lq locked" as a result of mutual friction.

In this Letter, we study the velocity statistics of quantum turbulence generated by a thermal counterflow on length scales between our experimental resolution ($\sim$1~$\mu$m) and the typical intervortex spacing ($\sim$0.1-1 mm) \cite{skrbek03}.  On such length scales the interactions of individual quantized vortices are important.  Specifically, quantized vortex reconnection \cite{schwarz85, koplik93, deWaele94, nazarenko03, bewley08}, where two vortices merge at a point, change topology by exchanging parts, and separate (Fig.\ 1 and \cite{movie1}), produces high, atypical velocities.  By analyzing the trajectories of micron-sized solid hydrogen tracers, we may compute both the velocity statistics of quantum turbulence, and identify and assess the effects of individual reconnection events.  Previous studies have shown that hydrogen tracers can be trapped on quantized vortices or, if not near a vortex, move with the normal fluid under the influence of Stokes drag \cite{poole05, bewley06, paoletti08}.

\begin{figure}
\begin{center}
\includegraphics[width=8.6cm]{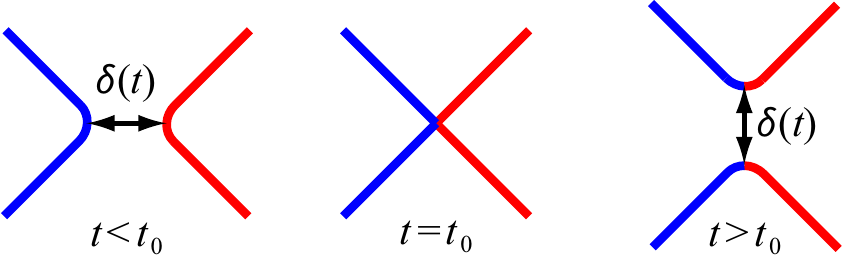}%
\caption[Reconnection schematic and example]{Depiction of one-dimensional topological vortices that reconnect by merging at the moment $t_0$ and then separate \cite{movie1}; the minimum separation distance is $\delta(t)$ with $\delta(t_0)=0$.}
\end{center}
\end{figure}

Our experiments are conducted in a cylindrical cryostat of 4.5~cm diameter using liquid $^4$He.  The long axis of the channel is vertical with four 1.5~cm windows separated by 90$^{\circ}$.  Particles are produced by injecting a mixture of 2\% H$_2$ and 98\% $^4$He into the liquid helium above the superfluid transition temperature (2.17~K) \cite{bewley06}.  The volume fraction of hydrogen is $\sim 10^{-8}-10^{-7}$, which results in each evident vortex having only a few trapped particles so that their effects may be neglected \cite{paoletti08}.  The fluid is then evaporatively cooled to the desired temperature in the range 1.70 K $<T<2.05$~K.  The hydrogen particles are illuminated by an argon-ion-laser sheet that is 8~mm tall and 100~$\mu$m wide.  Optical laser power varies between 2 and 6~W.  A video camera gathers 90$^{\circ}$ scattered light with a resolution of 16~$\mu$m per pixel at 60, 80, or 100 frames per second.

We study the velocity statistics in decaying quantum turbulence initiated by a reproducible thermal counterflow.  A spiral nichrome wire heater at the bottom of the channel 7.5~cm below the observation volume drives the counterflow.  The fixed heat flux varying from 0.064 to 0.17~W/cm$^2$ drives the system for approximately 5~s, after which it relaxes for about 10~s before the process is repeated.  A 2D particle-tracking algorithm with sub-pixel precision \cite{weeks} extracts single particle trajectories.

\begin{figure}
\begin{center}
\includegraphics[width=8.7cm]{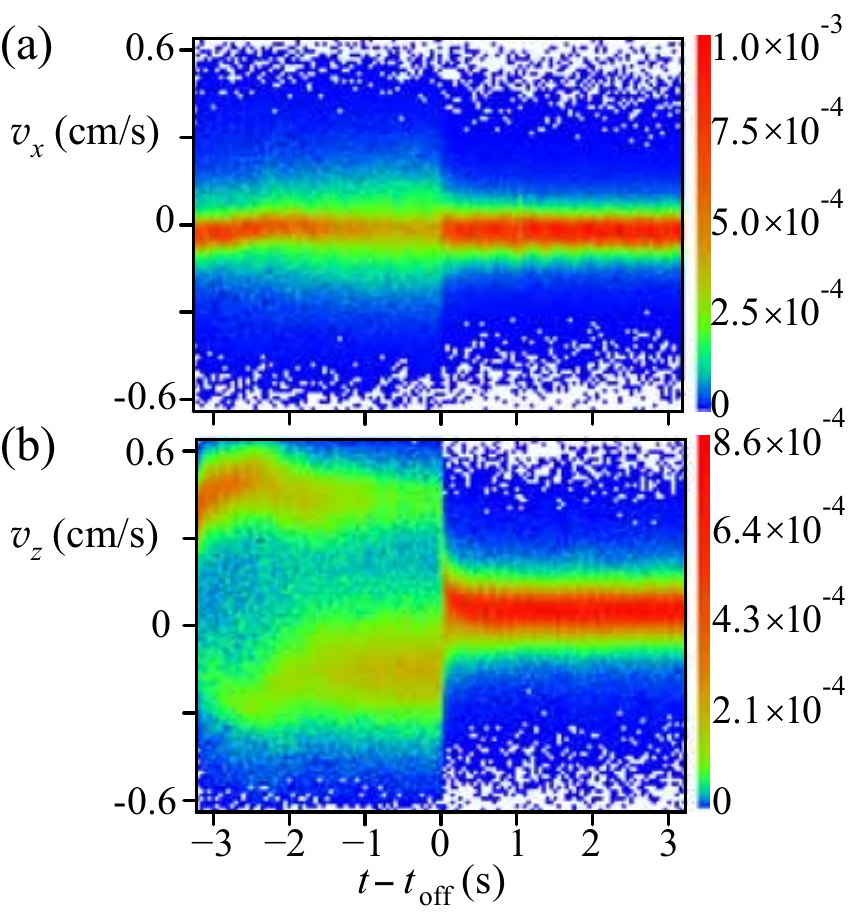}%
\caption[Two-dimensional velocity distributions]{Time-varying pulsed counterflow velocity distributions at $T=1.90$ K showing (a) $v_x$ and (b) $v_z$ for a portion of a heat pulse of 0.17 W/cm$^2$ with the heater turned off at $t=t_{\rm{off}}$ \cite{movie2}.  White denotes amplitudes with zero probability.}
\end{center}
\end{figure}

We characterize the resulting dynamics by analyzing the particle trajectories \cite{paoletti08}.  Time-varying distributions of the horizontal and vertical velocity components, $v_x$ and $v_z$, computed by forward differences, are shown in Fig.\ 2 for a typical thermal pulse \cite{movie2}.  The $v_x$ distributions are always peaked near zero. However, as predicted by the two-fluid model \cite{donnelly91}, the $v_z$ distributions exhibit a different behavior, since entropy injected by the heater is carried upward ($v_z>0$) by the motion of the normal fluid.  To conserve mass, the superfluid component moves downward opposing the normal fluid motion.  The bimodal $v_z$ distributions when the heater is on represent particles with $v_z>0$ moving upward with the normal fluid owing to Stokes drag while particles with $v_z<0$ are trapped in the vortex tangle that moves downward.  Once the heat pulse ends, the vertical velocities collapse to distributions peaked near zero.  The probability distribution functions (pdf) of velocity components derived from all trajectories for times after the heat pulse is turned off ($t>t_{\rm{off}}$) in Fig.\ 2 are shown in Fig.\ 3(a).  We focus on the tails of these distributions, which are composed of trajectories with high, atypical velocities; these we attribute to quantized vortex reconnection as explained below.

Near the reconnection moment, quantized vortices move with velocities much higher than the background flow \cite{movie1}.  These reconnections have been experimentally visualized using hydrogen particles \cite{bewley08} and studied numerically \cite{schwarz85, koplik93, deWaele94} and analytically \cite{nazarenko03}.  As discussed in some of these works, the minimum separation distance between reconnecting quantized vortices $\delta(t)$ (Fig.\ 1) evolves approximately as a square-root in time, i.e. $\delta(t)=A\sqrt{\kappa |t-t_0|}$, where $t_0$ is the reconnection moment and $A$ is a dimensionless factor of order unity.  Thus, for lengths between the vortex core radius and the typical intervortex spacing we expect the velocity to scale as
\begin{equation}
v(t)\propto |t-t_0|^{-1/2},
\label{vscaling}
\end{equation}
which grows much larger than typical fluid velocities when $t\rightarrow t_0$ (although cut off by the speed of sound).

In the pulsed counterflow experiments, a reconnection event is evidenced by a pair of nearby tracers rapidly approaching or separating.  Given the large number of possible particle pairs analyzed $(\sim 10^{10})$ an \it{ad hoc} \rm criterion is needed to select likely reconnection events.  We define a pair of particles $i$ and $j$ as marking reconnection at time $t$ if the separation $\delta_{ij}(t)=\vert\bf{r}\it_i(t)-\bf{r}\it_j(t)\vert$ satisfies
\begin{equation}
\label{deltaeqn}
\delta_{ij}(t\pm0.25 \rm{\;s})/\delta_{\it{ij}}(\it{t})>\rm{4},
\end{equation}
where $\bf{r}\it_i(t)$ is the two-dimensional projection of the position of particle $i$ at time $t$ and the plus (minus) sign denotes particles that separated after (approached before) an event, which we label as forward (reversed) events.  The duration of 0.25~s is chosen since greater times are dominated by boundary effects and the presence of other vortices. The criterion (\ref{deltaeqn}) excludes all but a fraction of possible pairs leaving $\sim 4 \times 10^{4}$ reconnection events.

The measured separations $\delta(t)$ for four typical forward events are shown in Fig.\ 4(a).  Solid line fits invoking a correction factor to the predicted scaling as
\begin{equation}
\delta^{\rm{fit}}(t)=A[\kappa(t-t_0)]^{1/2}[1+c(t-t_0)],
\label{deltafit}
\end{equation}
describe the data well \cite{powerlawfits}.  The fits minimize $\chi^2 \equiv n^{-1}\sum_{i=1}^n[(\delta^{\rm{fit}}_i-\delta_i)/\sigma]^2,$ where $i$ denotes the movie frame, $\sigma =4$ $\mu$m (0.25 pixels) is an estimate of the uncertainty of the particle positions, and $n=$ 15, 20, 25 for data collected at 60, 80, or 100 frames per second, respectively.  To fit reversed events, we use the same form (\ref{deltafit}) with $(t-t_0)$ replaced by $(t_0-t)$.  The distributions of $A$ and $c$ for forward (reversed) events, determined from fifty heat pulses, are shown in black (red) in Figs.\ 4(b) and (c).  These are calculated only from events with $\chi^2<4$; about 50\% of the pairs satisfying (\ref{deltaeqn}) meet this $\chi^2$ criterion.

\begin{figure}
\begin{center}
\includegraphics[width=8.7cm]{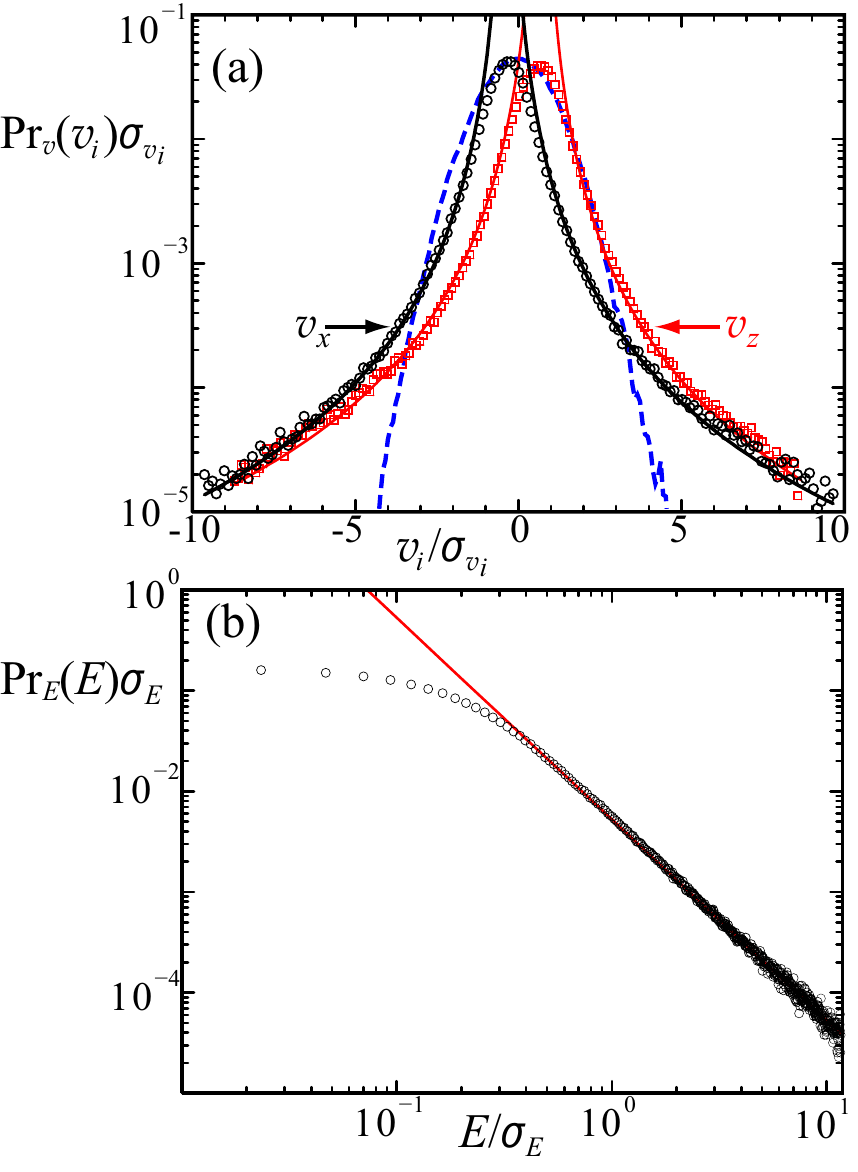}%
\caption[Statistics]{Local velocity and energy
statistics derived from the data in Fig.\ 2 for all particle trajectories with $t>t_{\rm{off}}$ (computed from over $1.1\times10^6$ values of velocity). All distributions are scaled to give unit variance using
$\sigma_{v_x}=0.066$, $\sigma_{v_z}=0.074$ cm/s or $\sigma_E=0.017$
(cm/s)$^2$.  (a) Probability distribution function of $v_x$ (black circles) and $v_z$ (red
squares).  The solid lines are fits to Pr$_v(v_i)=a\vert
v_i-\overline{v_{i}}\vert^{-3}$, where $i$ is either $x$ (black) or $z$ (red)
and $\overline{v_{i}}$ is the mean value of $v_i$.  For comparison the dashed (blue) line shows the distribution for classical turbulence in water \cite{zeff03} computed from over 10$^7$ velocity values.  The distribution is scaled using $\sigma_v=0.25$ cm/s, and with a peak value matched to the $v_x$ data.  The velocity statistics in water are close-to-Gaussian over five decades in probability.  (b) Probability distribution function of $E=(v_x^2+v_z^2)/2$ with a fit of the form Pr$_E(E)=aE^{-2}$ shown (in red).}
\end{center}
\end{figure}

\begin{figure}
\begin{center}
\includegraphics[width=8.7cm]{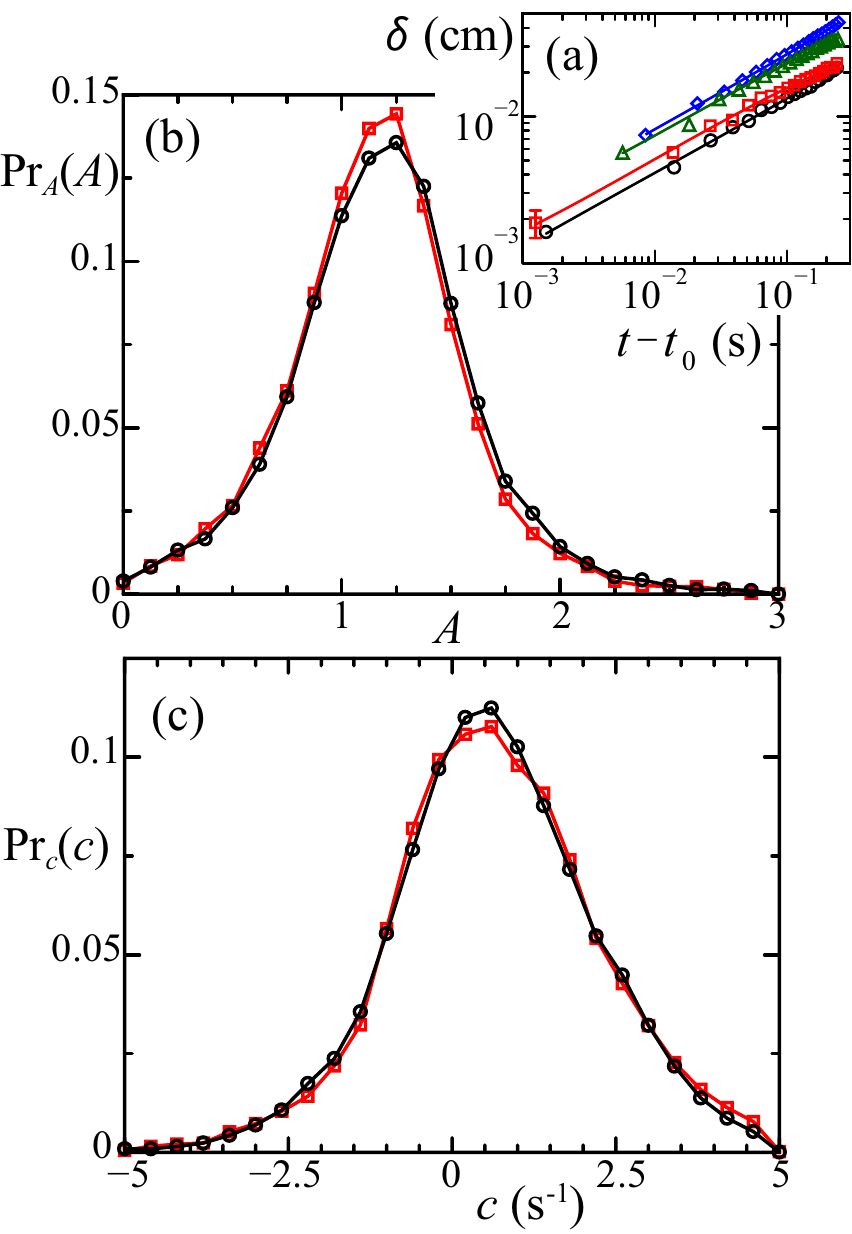}%
\caption[Scaling of reconnection]{Statistics of the reconnection fits.  (a) Four typical forward events.  Symbols denote the measured separation $\delta(t)$ of pairs of particles on reconnecting vortices with an example error bar $\sigma=4$ $\mu$m while solid lines show fits to (\ref{deltafit}). (b) Normalized distributions of the amplitude $A$ for 20,300 forward events (black circles) and 19,600 reversed events (red squares).  (c) Normalized distributions of the correction amplitude $c$ for the same 20,300 forward events (black circles) and 19,600 reversed events (red squares).}
\end{center}
\end{figure}

To model the pdf of the velocity derived from particle trajectories, we may use the transformation
\begin{equation}
\rm{Pr}\it_v(v)=\rm{Pr}\it _t[t(v)]|dt/dv|,
\end{equation}
where Pr$_v(v)dv$ is the probability of observing a velocity between $v$ and $v+dv$ at any time while Pr$_t(t)dt$ is the uniform
probability of taking a measurement at a time between $t$ and $t+dt$.  Hence, accepting the scaling relation (\ref{vscaling}), we predict for large $v$ (small $t$) the behavior
\begin{equation}
\rm{Pr}\it_v(v)\propto|dt/dv|\propto |v|^{\rm{-3}}.
\label{vfit}
\end{equation}

The $v_x$ and $v_z$ pdfs derived from all particle trajectories for $t>t_{\rm{off}}$ for the same pulse in Fig.\ 2 are shown in Fig.\ 3(a). The solid lines are fits to (\ref{vfit}) allowing for a mean flow.  To emphasize the distinction with classical turbulence, a velocity pdf from an oscillating-grid experiment in water \cite{zeff03} is also shown.  Evidently the velocity pdfs in superfluid helium differ drastically from the near-Gaussian velocity pdfs observed experimentally \cite{noullez97}, and in direct numerical simulations of homogenous and isotropic classical turbulence \cite{vincent91, gotoh02}.  One must note, however, that tracer particles in superfluid helium respond separately to the normal fluid and the quantized vortices (which are influenced by the normal fluid and superfluid); this is fundamentally different from that in water.

By the same argument, the tails of the pdf for the kinetic energy per unit mass $E=(v_x^2+v_z^2)/2$ will be dominated by reconnections.  Accepting the relation (\ref{vscaling}), we have $E(t)\propto\vert t-t_0\vert^{-1}$ and so for large $E$ we expect
\begin{equation}
\rm{Pr}\it_E(E)\propto\left|dt/dE\right|\propto E^{\rm{-2}}.
\label{Efit}
\end{equation}
The pdf of $E$ computed from the data in Fig.\ 3(a) (which includes all particle trajectories) is shown in Fig.\ 3(b).  The departure from the predicted power-law behavior for low velocities and energies may reasonably be attributed to effects from the boundaries and nearby vortices as well as to the background drift of the normal fluid.

In conclusion we have shown that the velocity statistics of quantum turbulence in superfluid $^4$He differ drastically from those for classical turbulence owing to the topological interactions of vortices that are different from those in classical fluids.  Previous studies argued that the interactions of magnetic field lines can cause the velocity statistics of magnetohydrodynamic (MHD) turbulence also to differ from classical turbulence.  The power-law tails in the distributions of electron energies observed in astrophysical settings (Fig.\ 3 in \cite{oieroset02} and Fig.\ 2 in \cite{holman03}) have been attributed to magnetic reconnection \cite{drake06}. Furthermore, theories for MHD turbulence propose that fractional diffusion may be the dominant transport mechanism \cite{del-castillo-negrete04}. Such diffusion is associated with power-law tails in velocity distribution functions. Since reconnection is a principal dissipative mechanism in superfluids near absolute zero, superfluid experiments might provide another \lq\lq laboratory" for studying strong MHD turbulence, as well as other systems exhibiting one-dimensional topological defects \cite{chuang91} such as liquid crystals, superconductors, Bose-Einstein condensates, and cosmic strings.

\begin{acknowledgments}
This work was supported by NSF-DMR, NASA, and CNAM at the University of Maryland.  We thank Gregory Bewley for past collaboration and Makoto Tsubota, Nigel Goldenfeld, Christopher Lobb, Marc Swisdak, and James Drake for helpful discussions.
\end{acknowledgments}


%\bibliography{QTvsCT}

\begin{thebibliography}{37}
\bibitem[{pow()}]{powerlawfits}
\bibinfo{note}{A model of the form $\delta(t)=A(t-t_0)^{\alpha}$ fits the data
  equally well and will be discussed in a future publication.}
\end{thebibliography}

\clearpage

\end{document}